\magnification=\magstep1
\baselineskip=16pt
\hfuzz=6pt

$ $

\rightline{Cambridge, November 13th 2025}

\vskip 1in

\centerline{\bf Thermodynamics + Natural Selection = Bayesian Inference}

\bigskip
\centerline{Seth Lloyd, MIT}

\vskip 1cm

\noindent{\it Abstract:} Consider a population of organisms that 
harvest free energy from their environment to reproduce.   This paper shows 
that if the organisms' reproductive rates are proportional to the amount of physical
free energy that they can convert into reproductive work, then the implicit 
probabilities that the organisms assign to environmental states are
updated according to Bayes' rule.

\vskip 1cm

Free energy is energy that is available to do work [1-7].   Living organisms harvest
free energy from their environment and use that free energy to perform 
biochemical work.       
The goal of this paper is to make mathematically rigorous the
intuition that
a system that has a better model of its environment ought
to be able to harvest more free energy, to perform more biochemical
work more efficiently, and to reproduce more.  
A photosynthetic organism that produces a pigment that is
sensitive to light at a new frequency, or animal that can produce
an enzyme that allows it to metabolize a previously indigestible
molecule, can convert more free energy to work 
and gains a potential reproductive advantage over its competitors. 

Consider a population of self-reproducing organisms that 
interact with a non-equilibrium subsystem of the environment together with
a thermal bath to convert free energy into work.  Although the organisms do not explicitly estimate the probabilities  
of the states of the non-equilibrium system, we show that -- based on the actions an organism takes to convert free energy to work -- each organism possesses an implicit model of the physical probabilities.   We show that an organism
with a better implicit model can convert more free energy to work:
the free energy lost due to an inaccurate model is proportional to
relative entropy between the actual, physical probabilities and the 
probabilities as predicted by the model.  But this relative entropy
can be shown to be
the {\it same} quantity that governs the growth rate of a population
of models that perform Bayesian inference [3-7].  As a consequence, if
the growth rate of an organism/model is proportional to the 
amount of free energy the organism can convert to reproductive work,
the population as a whole effectively performs Bayesian inference. 

Organisms that perform
Bayesian inference maximize the conversion of free energy into work.
Organisms that maximize the conversion of free energy into work perform Bayesian inference.

\bigskip\noindent{\it 1. Free energy and dissipation}

Consider a system whose states $x$ have energies $E_x$.    If the probabilities for the states $x$ are $p(x)$, then the amount of work that can be extracted from the system via
adjusting the energies $E_x$ and by interacting with a bath at temperature $T$ is equal to the difference between the system's non-equilibrium free
energy 
$$F = \sum_x p(x) E_x
-T (-\sum_x p(x) \ln p(x)) = E -TS \eqno(1)$$ and its
equilibrium free energy
$$F_{eq} = \sum_x p_{th}(x) E_x
-T (-\sum_x p_{th}(x) \ln p_{th}(x)),\eqno(2)$$
where 
$p_{th}(E_x) = 
(1/Z(T)) e^{-E_x/T}$ are the Boltzmann-Gibbs thermal probabilities and
$Z(T) = -\sum_x e^{-E_x/T}$ is the partition function (we work in units where Boltzmann's constant is $1$).

The work that can be extracted from a sequence of adiabatic/entropy-preserving steps and isothermal steps (as in the Carnot process)  is then [1] 
$$W = F - F_{eq} = 
T D(p||p_{th})\eqno(3)$$
where 
$D(p||p_{th}) = -\sum_x
p(x)\ln p_{th}(x)/p(x)$ is
the relative entropy/Kullbach-Leibler distance between
$p(x)$ and $p_{th}(x)$. 
For example, the work $W$ can be extracted by first performing an adiabatic/isentropic quench that
changes the energies $E_x$ to
$E'_x = - T \ln p(x)$, so that the new energy for the state $x$ is equal to its
thermal probability at temperature $T$.   We then isothermally restore the energies to their original values [1].    This process extracts the work given by equation (3).   
 
The adiabatic/isentropic steps of the work extraction process are energy and entropy conserving.    However, extra dissipation arises when during the isothermal step if one changes the energies to the `wrong' values $E''_x \neq - T \ln p(x)$.   The choice of energies $E''$ defines {\it implicit} probabilities $q(x) = 
-\ln E''_x/T$: $q(x)$ are the probabilities which -- if they were equal to $p(x)$ -- would result in the maximal conversion of free energy to work.   

When the implicit probabilities $q(x)$ are not equal to the physical probabilities $p(x)$ the free energy dissipated when the -- now non-thermal -- state is put into contact with the thermal bath and comes to equilibrium is 
$$ \Delta F_{lost} = F_q - F_p 
= \sum_x p(x) E''_x  -TS  - ( \sum_x p(x) E'_x -TS) =  T  D(p \| q). \eqno(4)$$
That is, the loss of free energy/work arising from an incorrect estimate of the actual physical probabilities when putting the system in contact with the thermal bath is equal the temperature times the relative entropy between the physical probabilities and the implicit probabilities.

\bigskip

\noindent{\it (2) Natural selection and Bayesian inference}

Consider a population of reproducing organisms.   Each organism has access to a non-equilibrium subsystem of the environment with states $x$, physical probabilities $p(x)$, and energies $E_x$ as above.  The organism uses a sequence of adiabatic and isothermal steps to convert free energy into work: each organism 
can interact with the system by changing the energies of its states,
and by putting it in contact with a thermal bath at temperature $T$.
After the interaction is completed, the energies of the system's states
are returned to their initial, non-interacting values. 

Assume that the rate of reproduction of an  organism that adopts a particular free energy conversion strategy is proportional to the amount of free energy it is able to convert into work.  The free energy extraction process is repeated multiple times.  We now show that the population of organisms as a whole performs Bayesian inference, in the sense that the implicit probabilities $q(x)$ that the population assigns to the states $x$ are updated according to Bayes's rule [3-7].  

First, review the dynamics of a such a population of organisms under Bayes rule, {\it without} any thermodynamic considerations.  
Let $q(m)$ be the fraction of
of the self-replicating population    
that possess a probabilistic model
$m$ that assigns probabilities $q(x|m)$ 
to the states $x$ of the environment during the free energy conversion process as described above.  Let $q_0(m)$ 
the prior probability of the model $m$.    
When the environment is found to have state $x$, the prior probability 
$q_0(m)$ is updated to its posterior probability $q(m|x)$ according to 
Bayes's rule [3-7]:
$$ { q(m|x) \over q_0(m)} = { q(x|m) \over q(x) }, \eqno(5)$$
where $q(x) = \sum_m q_0(m) q(x|m)$ is the ensemble prediction
for the state $x$, averaged over the ensemble.   

Now look at the expected growth rates for different models over multiple rounds of Bayesian inference.
Taking the logarithm of equation (2), and averaging the over
the physical probabilities $p(x)$,
we find that 
$m$'s average growth rate is 
$$
\sum_x p(x) \ln q(m|x)/q_0(m) = \sum_x p(x) \ln q(x|m)/q(x) = 
- D(p(x) \| q(x|m) ) + D( p(x) \| q(x) ). \eqno(6)$$
That is, the average growth rate of model $m$ is given by the
relative entropy between the true probabilities
$p(x)$ and the ensemble prediction $q(x)$,
minus the relative entropy between the true probabilities
and the probabilities $q(x|m)$ predicted by the model $m$.
Models whose predicted probabilities $q(x|m)$ are closer to the 
actual physical probabilities $p(x)$, as measured by smaller
relative entropy, have a larger average growth rate, and will in time come to dominate the population.

Comparing equations (4) and (6), we see that 
the penalty in growth rate for organisms with model $m$ 
is equal the one over the temperature times the extra dissipation in free energy that such 
an organism suffers from assigning probabilities $q(x|m)$ that do not match the physical probabilities $p(x)$: both are equal to the relative entropy between the physical probabilities and model's probabilities.   More precisely,
the growth rate of organism with model $m$ minus the average growth rate over the population, is equal to the free energy converted to work by the organism minus the average free energy converted to work by the population as a whole, divided by the temperature.

This equality is our main result: an organism's efficiency in converting free energy to work, divided by the temperature, is equal the growth rate of the organism compared with the overall population when the population is performing Bayesian inference. 

\bigskip\noindent{\it 3. Remarks}

\bigskip\noindent (I) The results above were derived for classical probabilities.   The quantum
version can be obtained simply by substituting in the density
matrix $\rho$ for the non-equilibrium system, and considering models that
give predicted density matrices $\sigma$.  All the results still hold,
but now in terms of the quantum relative entropy $D(\rho\| \sigma) - = {\rm tr} \rho ( \sigma - \rho)$ and thermal Hamiltonians of the form $ H_{th} =  -T \ln \rho$.   In the quantum case, in addition to 
changing the energies of the system to match the thermal probabilities,
the organism must also match the energy eigenstates $|E_j\rangle$ of $H_{th}$ to 
the eigenstates of $\rho$.   Let $\tilde \rho = \sum_j |E_j\rangle \langle E_j|  \rho |E_j\rangle \langle E_j|$.   
An inaccurate estimate of the energy eigenstates of the quantum
system will incur additional dissipation $T D(\rho \|\tilde\rho)$ due to decoherence during
the thermal quench step above.

\bigskip\noindent (II) The results here concern the physical free energy, defined in terms of $D(p\|q)$, not Friston's variational 
free energy [8], which is defined in terms of the relative entropy $ D(q \| p)$ (see, e.g., [6]).   Although the two relative entropies have the same fixed point, they are not in general equal.
    Physical free energy, not variational free 
energy, is the quantity that governs the ability of an organism to convert
free energy to work.

\bigskip\noindent(III) The derivations above assumed that the environment has no memory.
The results are easily extended to correlated sequences of states governed by physical probabilities $p(x_1 \ldots x_n)$ and estimated probabilities $q(x_1 x_n\ldots)$ over multiple rounds of work extraction, where $x_j$ is the state of the subsystem at the $j$'th round.  As long as $p(x_1 \ldots x_n)$ is stationary and ergodic, then both the amount of free energy converted to work and the growth rate of organisms/models performing Bayesian inference are governed by the relative entropy $D(p\|q)$ over sequences of states.  

\bigskip\noindent{\it 4. Discussion}

Free energy and Bayesian inference are both intimately related to relative entropy.   In this paper, we showed that the extra energy dissipation associated with having an incorrect model of environmental probabilities is proportional to the temperature times the relative entropy between the actual physical probabilities and the model probabilities.   But the same relative entropy also governs the penalty in growth rate for a model in a population performing Bayesian inference.    Consequently, if the reproductive rate of an organism is proportional to the amount of free energy it can convert into work, a population of organism evolving under natural selection performs Bayesian inference.
Thermodynamics  + Natural selection = Bayesian inference.

\vfill\eject
\noindent{\it Acknowledgements:}  The author thanks Michele Reilly, Sam Gershman,
and Jordan Horowitz for helpful discussions.
This material is based upon work supported by, or in part by, the U. S. Army Research Laboratory and the U. S. Army Research Office under contract/grant number W911NF2310255, and by DoE under contract, DE-SC0012704.

\vskip 1cm
\noindent{\it References:}

\smallskip\noindent [1]
J.M.R. Parrondo, J.M. Horowitz, T. Sagawa, `Thermodynamics of
Information,' {\it Nat. Phys.} {\bf 11}, 131 (2015).

\smallskip\noindent [2]
A. Kolchinsky, I. Marvian, C. Gokler, Z.-W. Liu, P. Shor, O. Shtanko,
K. Thompson, D. Wolpert, S. Lloyd, `Maximizing free energy gain,'
{\it Entropy} {\bf 27} 91 (2025); arXiv: 1705.00041 (2017).

\smallskip\noindent [3]
Marc Harper, `The replicator equation as an inference dynamic,'
arXiv: 0911.1763v3 (2010).

\smallskip\noindent [4]
J.C. Baez and B.S. Pollard, `Relative entropy in biological systems,'
{\it Entropy} {\bf 18}, 46 (2016).

\smallskip\noindent [5]
Martin Nowak, `Evolutionary Dynamics: Exploring the Equations of Life,'
Belknap Press of Harvard University Press, Cambridge 2006.

\smallskip\noindent [6]
S.J. Gershman, `What does the free energy principle tell us about the 
brain?'  arXiv: 1901.07945 (2019).

\smallskip\noindent [7]
T.M. Cover, J.A. Thomas, {\it Elements of Information Theory,} 2nd
edition, Wiley 2006

\smallskip\noindent [8] K. Friston {\it Nat. Rev. Neuro.} {\bf 11},
127-138 (2010).

\vfill\eject\end

\bigskip\noindent{\it (3) Free energy and extractable work}

Consider an organism that attempts to extract work or storable energy from
a non-equilibrium subsystem of its environment.    The organism 
can interact with the system by changing the energies of its states,
and by putting it in contact with a thermal bath at temperature $T$.
After the interaction is completed, the energies of the system's states
are returned to their initial, non-interacting values.

As will be seen below, a system that uses
model $m$ to extract work from the environment in this manner effectively
possesses an implicit estimate $q(x|m)$ of the actual probabilities $p(x)$ of
physical states $x$.  Any mismatch 
between estimated probabilities $q$ and the physical probabilities $p$
necessarily results in dissipation and loss of free energy.   
Below, we will use the physics of free energy harvesting to prove the 
following
\smallskip\noindent
{\it Theorem:} The amount of free energy lost because an incorrect estimate of
physical probabilities is 
$$ \Delta F_{lost} = T D(p(x) \| q(x|m)). \eqno(4)$$
\smallskip
That is, systems that maximize the harvested free energy are
those that minimize the relative entropy betwen the physical probabilities
and the estimated probabilities.   These are exactly the same systems
that come to dominate a population performing Bayesian inference.

More precisely, as we now prove in detail,
if the average growth rate of an organism is proportional
to the free energy it converts to work, then the population of
reproducing, free-energy harvesting organisms as a whole is effectively
performing Bayesian inference.   

\bigskip\noindent{\it (3.1) Non-equilibrium free energy}

Suppose that the system that the organisms are using to harvest free energy 
and to convert it to work has states $x$, with energies $E(x)$, and initial 
probabilities $p(x)$.    The non-equilibrium free energy of the system is
defined to be [1-2]
$$ F = E - TS = \sum_x p(x) E(x) + T \sum_x p(x) \ln p(x).  \eqno(5)$$
The amount of work that can be extracted from the system by interacting
with it to change its energy, and by putting it into contact with the thermal
bath is [1]
$$W_{ex} = F - F_{eq}, \eqno(6)$$
where $F_{eq}$ is the equilibrium free energy, defined relative to the
thermal probabilities    
$p_{th}(x) = e^{-E(x)/T}/Z$, where $Z = \sum_x
e^{-E(x)/T}$ (we use units in which Boltzmann's constant equals 1, so 
that temperature is measured in joules rather than degrees).    
When the system is already at thermal equilibrium, no work can be 
extracted by having it interact with the thermal bath.   

This work can be extracted by performing the following
steps [1]:

\smallskip\noindent (1) {\it Isentropic quench:} 
With the subsystem out 
of contact with the bath, change the energies of the states 
to take the values $E_p(x) = - T \ln p(x)$.   The probabilities
$p(x)$ and the entropy remain unchanged.  

\smallskip\noindent (2) {\it Thermal contact:}
Put the subsystem in contact with the bath.
Because the probabilities $p(x)$ are now the thermal probabilites with
respect to the new energies $E_p(x)$, the system is in equilibrium and no
dissipation takes place.

\smallskip\noindent (3) {\it Isothermal restoration:} With the system in contact
with the bath, gradually restore the energies to their original values:
$E_p(x) \rightarrow E(x)$.    At the end of the isothermal restoration the 
subsystem is at thermal equilibrium at temperature $T$.  
Heat flows between subsystem and bath, but
if the isothermal dynamics is sufficiently slow, entropy increase can 
be made vanishingly small.      

\smallskip
At the end of this process, the total work extracted by the organism that
is interacting with the system by changing its energies is easily verified to be
$F-F_{eq}$.

\bigskip\noindent{\it (3.2) The cost of ignorance}

The method of [1] requires knowledge of the true physical probabilities
$p(x)$.   Suppose that instead the system using model $m$ possesses
an inexact estimate $q(x|m)$
of these probabilities.   We now show that the energetic cost of 
having an incorrect estimate of the true probabilities is governed
by the relative entropy/Kullbach-Leibler distance
between the correct and the estimated probabilities, that is,
by the {\it same} quantity that governs the updating
of probabilities/populations via Bayesian inference.

Suppose that during the adiabatic quench, the
system trying to extract the work sets the
energies to values that are determined by the estimated
probabilities $q(x|m)$ rather than by the physical probabilities $p(x)$:
$$ E_m(x) = - T \ln q(x|m) \neq - T \ln p(x) = E_p(x).\eqno(6)$$ 
The probabilities $p(x)$ are no longer the thermal probabilites
$q_{th} \propto e^{ - E_m(x)/T }$ for the subsystem, 
and when the subsystem is put in contact with
the bath at temperature $T$, it is not at equilibrium. 
Energy is now dissipated as the subsystem re-equilibrates.   

The lost free energy is equal to
the difference between the non-equilibrium free energy defined with
respect to the actual probabilities $p(x)$ and the new energies $E_m(x)$, 
and the equilibrium free energy with those energies:
$$ \eqalign{ \Delta F_{lost} &=  \sum_x p(x) E_m(x) + T\sum_x p(x) \ln p(x)
-\sum_x q_{th}(x) E_m(x) - T \sum_x q_{th}(x) \ln q_{th}(x) \cr
&= - T \sum_x p(x) \ln q(x|m)/p(x)\cr
& = T D(p(x) \| q(x|m)). \cr}
\eqno(7)$$  
This is just the theorem stated above (equation 4):
the energy dissipated by applying the `wrong' energies, based on
the estimated probabilities rather than the physical probabilities,
is equal to the temperature of the bath times
the relative entropy/Kullbach-Leibler distance between the
physical probabilities and the estimated probabilities. 

\bigskip\noindent{\it 3.2 Free energy harvesting: Discussion}

We emphasize that the relationship
between relative entropy and free energy dissipated, equation (4), 
holds generally for any method that converts non-equilibrium free energy
into work/stored energy by interaction with a bath at temperature
$T$.  If the energies of the
system are to be restored to their original
values at the end of the process, then   
to extract work, it is necessary at some point of the process
to put the system into contact with the
bath.   At that point, if the estimated probabilities are not
equal to the true probabilities, the energy dissipated is
equal to the temperature times the relative entropy between
the true and estimated probabilities.   

Relative entropy governs the amount of dissipation
that occurs because of contact with the thermal bath, which
in our simplified model is the only source of entropy increase --
we have assumed that the interaction on its own is adiabatic/isentropic: it
only changes the energies of states and not their probabilities. 
If the interaction is not isentropic then our theorem provides
a lower bound on the energy dissipated.

\bigskip\noindent{\it (4) Comparison}

Equation (3) 
shows that under Bayesian inference,
the average increase in log probability for a model $m$ with
an estimate $q(x|m)$ for the true probabilities $p(x)$ is governed
by the relative entropy between $p(x)$ and $q(x|m)$. The closer
the estimates of the model to the true probabilities, the smaller
the relative entropy between $p$ and $q$, 
the greater the model's average growth rate.  

Theorem (1) 
shows that for an organism that uses model $m$ to harvest free energy from
its environment, the amount of dissipation caused by having a `wrong' estimate
$q(x|m)$ for the true physical probabilities for states $p(x)$ is 
governed by the relative entropy between $p(x)$ and $q(x|m)$.  The 
the closer the estimated probabilities are to the true 
probabilities, the smaller the relative entropy between $p$ and $q$, 
the greater the amount of free energy that can be harvested.  

We now combine theorem (1) for free energy harvesting and equation (3)
for Bayesian inference to
compare the amount of additional free energy harvested by an organism
that uses model $m$ to the amount harvested by an organism that uses model
$m'$: 
$$ \Delta F_{lost}(m) - \Delta F_{lost}(m') 
= T ( D(p(x) \| q(x|m) -  D(p(x) \| q(x|m'))
 = T (\bar \lambda(m) - \bar\lambda(m')), \eqno(8) $$
where $\bar\lambda(m)$, defined in equation (3) above, is the average 
growth rate
of the model $m$ under Bayesian inference. 
That is, the added efficiency in free energy harvesting obtained by
model $m$ over $m'$ by having a better estimate of the physical
probabilities, is equal to the temperature times the 
advantage in average rate of growth obtained by model $m$ over $m'$ by having a better
estimate of the physical probabilities. 

\bigskip\noindent
Stated in less technical terms: A population whose average
growth rates are proportional to the amount
of environmental free energy that can be converted into reproductive work 
will evolve via Bayesian inference, and will become dominated by
implicit models whose predictions of
environmental probabilities are closer, 
in the sense of relative entropy, to the 
true physical probabilities of the environment.    
The advantage in free energy converted is equal to the temperature
times the advantage in growth rate. 

\bigskip\noindent{\it 4. Discussion}

Before concluding, several notes are in order.

\smallskip\noindent (I)
The results above were derived for classical probabilities.   The quantum
version can be obtained simply by substituting in $\rho$ for the density
matrix for the non-equilibrium system, and considering models that
give predicted density matrices $\sigma$.  All the results still hold,
but now in terms of the quantum relative entropy $D(\rho\|\sigma)
= - {\rm tr} \rho \log \sigma + {\rm tr} \rho \log \rho$ between
the physical and estimated density matrices, and thermal Hamiltonians
of the form $ H_{th} =  -T \log \rho$.   In the quantum case, in addition to 
changing the energies of the system to match the thermal probabilities,
the organism must also match the energy eigenstates of $H_{th}$ to 
the eigenstates of $\rho$:   
an inaccurate estimate of the energy eigenstates if the quantum
system will incur additional dissipation due to decoherence during
the thermal quench step above.

\smallskip\noindent (II) The results here concern the physical free energy, not Friston's variational 
free energy [8].   Phrased in the notation above, variational free energy
is defined in terms of $T D(q(x) \| p(x))$, 
where $q(x)$ are the estimated/implicit
probabilities and $p(x)$ are the physical probabilities.  By contrast, 
the amount of physical energy that free energy can be converted into to work 
is determed by 
$T D(p(x) \| q(x))$, which in general $\neq T D(q(x) \| p(x))$ due to the
asymmetry of relative entropy.    Physical free energy, not variational free 
energy, is the quantity that governs the ability of an organism to convert
free energy to work.
A succinct discussion of the 
difference between Bayesian optimization and the variational free
energy principle can be found in [6].

\smallskip\noindent (III) The optimization problem faced by the 
population is convex in the implicit probabilities and linear in the
estimated energies.   The optimum lies in the interior of the 
probability/energy spaces and so the search problem
possesses no local optima.  A population that can freely explore
the space of implicit probabilities/estimated energies will evolve to the 
Bayesian/free energy harvesting
optimum, where implicit predicted probabilities and estimated energies
are equal to the physical probabilities and energies.  

\bigskip\noindent{\it Conclusion:}

The connection between physical free energy and relative entropy, and the
connection between Bayesian inference and relative entropy, are well known.
This paper `closes the loop' by showing that when average growth rates within
a reproducing population are proportional to the amount of physical free energy
that an organism can convert to work, the population effectively
performs Bayesian inference on its non-equilibrium environment.

\vfill
\noindent{\it Acknowledgements:} This work was supported by ARO, AFOSR, and
DARPA.    The author thanks Sam Gershman,
Michele Reilly, and Jordan Horowitz for helpful discussions.

\vfill\eject
\noindent{\it References:}

\smallskip\noindent [1]
J.M.R. Parrondo, J.M. Horowitz, T. Sagawa, `Thermodynamics of
Information,' {\it Nat. Phys.} {\bf 11}, 131 (2015).

\smallskip\noindent [2]
A. Kolchinsky, I. Marvian, C. Gokler, Z.-W. Liu, P. Shor, O. Shtanko,
K. Thompson, D. Wolpert, S. Lloyd, `Maximizing free energy gain,'
{\it Entropy} {\bf 27} 91 (2025); arXiv: 1705.00041 (2017).

\smallskip\noindent [3]
Marc Harper, `The replicator equation as an inference dynamic,'
arXiv: 0911.1763v3 (2010).

\smallskip\noindent [4]
J.C. Baez and B.S. Pollard, `Relative entropy in biological systems,'
{\it Entropy} {\bf 18}, 46 (2016).

\smallskip\noindent [5]
Martin Nowak, `Evolutionary Dynamics: Exploring the Equations of Life,'
Belknap Press of Harvard University Press, Cambridge 2006.

\smallskip\noindent [6]
S.J. Gershman, `What does the free energy principle tell us about the 
brain?'  arXiv: 1901.07945 (2019).

\smallskip\noindent [7]
T.M. Cover, J.A. Thomas, {\it Elements of Information Theory,} 2nd
edition, Wiley 2006

\smallskip\noindent [8] K. Friston {\it Nat. Rev. Neuro.} {\bf 11},
127-138 (2010).

\vfill\eject\end